\documentclass[aps,showpacs,preprintnumbers,nofootinbib,amsmath,amssymb]{revtex4-1}
\usepackage{amssymb}
\usepackage{slashed,color,amsmath,amssymb,mathrsfs}
\usepackage{graphicx}
\usepackage{ulem}
\usepackage{hyperref}
\usepackage{longtable}
\usepackage{array}
\usepackage{bm}

\allowdisplaybreaks

\begin{document}
\preprint{}
\title{Uniqueness of the Observable Leaving Redundant Imprints in the Environment in the Context of Quantum Darwinism}
\author{Hui-Feng Fu}
\email{huifengfu@jlu.edu.cn}
\affiliation{Center for Theoretical Physics, College of Physics, Jilin University,
Changchun 130012, P. R. China}

\begin{abstract}
{
In quantum Darwinism, the pointer observable of a system leaves redundant imprints in its environment after decoherence. Each imprint is recorded in a fraction of the environment, which  identifies a particular partition of the environment. An ambiguity situation may occur when another observable noncommuting to the pointer observable also leaves redundant imprints with respect to another partition of the environment. We study this problem based on a uniqueness theorem we proved. We find that within a particular subset of all possible partitions of the environment, the observable of the system leaving redundant and nondegenerately recorded imprints in the environment is unique. And, in a typical situation, the partitions outside this particular subset have no physical significance.
}
\end{abstract}
 \maketitle
\section{Introduction}

The environment of macroscopic objects plays an important role in the quantum-to-classical transition problem. On one hand, the environment of a macroscopic system can destroy interference between the states of the system through the (unavoidable) system-environment interaction. This is known as decoherence or environment-induced decoherence~\cite{Dec0,Dec1}. On the other hand, based on the stability argument, environment selects a set of states (referred to as preferred states or pointer states) that are stable in spite of the presence of the system-environment interaction. Here ``stable" means that if the initial state is a pointer state, it remains unentangled with its environment. Which set of the states are selected is essentially determined by the system-environment interaction. This is known as environment-induced superselection or einselection for short~\cite{Ensel0,Ensel1}. Typically, the pointer states are just the classical states we perceive in our daily lives, such as the approximate eigenstates of the center-of-mass position of a macroscopic object. In ideal situation, pointer states form a basis of the system's Hilbert space and are eigenstates of the so-called pointer observables. For more details of decoherence theory, we refer the readers to the reviews~\cite{Dec_rev-1,Dec_rev0,Dec_rev1,Dec_rev2} and references therein.

Another important role played by environment has been recognized in quantum Darwinism~\cite{QDar0,Env0,QDar1,QDar2,QDar3,QDar4}. Through decoherence process, selective information of the system spreads redundantly into the environment. According to quantum no-cloning theorem~\cite{Noclone}, the environment cannot copy the (unknown) entire quantum state of the system, so only the information of selective observables can be copied. The environment of a macroscopic object typically consists of numerous subsystems, such as photons, molecules, etc.. So after decoherence and proliferation of selective information, the information would have many copies, and each of the copies is recorded in a fraction of the environment. The importance of the existence of redundant records of selective information in the environment is that it allows many observers to observe the selected properties of the system independently without disturbing it. This is essential for the ``classical world" to be objective. Here, a property of a system is objective if it is (1) simultaneously accessible to many independent observers, (2) who can find out what it is without prior knowledge, and (3) who can arrive at a consensus about it~\cite{Env0}. Quantum Darwinism has been studied in various dynamical models~\cite{QDyn0,QDyn1,QDyn2,QDyn3,QDyn4,QDyn5,QDyn6}. In recent years, another idea that can explain the information redundancy and the emergence of the objectivity from quantum world, known as the spectrum broadcast structure framework, has also been proposed and studied~\cite{SBS0,SBS1,SBS2,SBS3}.

Quantum Darwinism tells us that selective observables of the system can be recorded redundantly in the environment, which is actually a selection rule. In decoherence theory, pointer observables are selected according to the environment-induced superselection rule. The two selection rules should coincide. It has been shown in Refs.~\cite{QDar0,Env0} that the only observables that can leave sufficiently many imprints in the environment are the pointer observables. However, if the number of the imprints are not sufficiently large (the meaning of ``sufficiently" will be clear shortly), an ambiguity situation may occur. For example, consider a four-dimensional system interacting with its environment, which consists of four two-dimensional elementary subsystems labeled as $\alpha, \beta, \gamma, \delta$. Suppose $\hat{\Lambda}\equiv\sum_{i=1}^4 \lambda_i |\phi_i^S\rangle\langle\phi_i^S|$ is the pointer observable, where $|\phi_i^S\rangle$ ($i=1,2,3,4$) form an orthonormal basis in the Hilbert space of the system, and the eigenvalues $\lambda_i$'s are all different. If, after an ideal decoherence process, the state of the whole system (including environment) becomes
\begin{eqnarray}\label{Am_ex0}
|\Psi\rangle&=&\frac{1}{\sqrt{4}}\sum_{j=1}^4 |\phi_j^S\rangle|\phi_j^A\rangle|\phi_j^B\rangle \notag \\
&=&\frac{1}{4}\Big\{|\phi_1^S\rangle(|00\rangle+|11\rangle)_A(|00\rangle+|11\rangle)_B+|\phi_2^S\rangle(|01\rangle+|10\rangle)_A(|01\rangle+|10\rangle)_B  \notag\\
&&+|\phi_3^S\rangle(-i|01\rangle+i|10\rangle)_A(-i|01\rangle+i|10\rangle)_B+|\phi_4^S\rangle(|00\rangle-|11\rangle)_A(|00\rangle-|11\rangle)_B \Big\},
\end{eqnarray}
where the subscripts $A$, $B$ represent fractions of the environment: $A=\{\alpha, \beta\}$, $B=\{\gamma, \delta\}$, and we have used the notation of qubits to express the environmental subsystems. (We are considering perfect correlations here.) Then the whole information of the pointer observable $\hat{\Lambda}$ has been recorded in the environment with two copies in the fractions $A$ and $B$, i.e., the pointer observable leaves redundant imprints in the environment.
However, noticing that the environment can be partitioned in different ways, the state $|\Psi\rangle$ can as well be expressed in another form, with respect to (w.r.t.) another partition of the environment: $a=\{\alpha, \gamma\}$, $b=\{\beta, \delta\}$, as
\begin{eqnarray}\label{Am_ex1}
|\Psi\rangle&=&\frac{1}{\sqrt{4}}\sum_{j=1}^4 |\psi_j^S\rangle|\psi_j^a\rangle|\psi_j^b\rangle \notag\\
&=&\frac{1}{4}\Big\{|\psi_1^S\rangle(|00\rangle+|11\rangle)_a(|00\rangle+|11\rangle)_b+|\psi_2^S\rangle(|00\rangle-|11\rangle)_a(|00\rangle-|11\rangle)_b  \notag\\
&&+|\psi_3^S\rangle(|01\rangle+|10\rangle)_a(|01\rangle+|10\rangle)_b+|\psi_4^S\rangle(|01\rangle-|10\rangle)_a(|01\rangle-|10\rangle)_b \Big\},
\end{eqnarray}
where
\begin{eqnarray}
|\psi_1^S\rangle = \frac{1}{2}(|\phi_1^S\rangle+|\phi_2^S\rangle-|\phi_3^S\rangle+|\phi_4^S\rangle), &~~~~&
|\psi_2^S\rangle = \frac{1}{2}(|\phi_1^S\rangle-|\phi_2^S\rangle+|\phi_3^S\rangle+|\phi_4^S\rangle), \notag\\
|\psi_3^S\rangle = \frac{1}{2}(|\phi_1^S\rangle+|\phi_2^S\rangle+|\phi_3^S\rangle-|\phi_4^S\rangle), &~~~~&
|\psi_4^S\rangle = \frac{1}{2}(|\phi_1^S\rangle-|\phi_2^S\rangle-|\phi_3^S\rangle-|\phi_4^S\rangle). \notag
\end{eqnarray}
$|\psi_i^S\rangle$ ($i=1,2,3,4$) form another orthonormal set, and could be eigenstates of another observable which does not commute with the pointer observable $\hat{\Lambda}$. So another observable non-commuting to the pointer observable also leaves redundant imprints in the environment according to Eq. (\ref{Am_ex1}).

The ambiguity of this type has been studied in Refs.~\cite{Env0,Loc}. In Ref.~\cite{Env0}, the authors proved that as long as the redundancies of two observables $R_0(\hat{X_1})$ and $R_0(\hat{X_2})$ (where $R_0(\hat{X})$ is, roughly speaking, the number of the copies recording the information about $\hat{X}$ in the environment) satisfy $R_0(\hat{X_1})R_0(\hat{X_2})> N$ where $N$ is the number of the environmental subsystems, the ambiguity will not occur. So, they set a requirement that the redundancy should be sufficiently large, i.e., $R_0(\hat{X})>\sqrt{N}$. The example given above does not satisfy this requirement. In Ref. \cite{Loc}, the author proposed a criterion which identifies a unique decomposition of the state of the whole system. Due to the similarity of the work of Ref.~\cite{Loc} and this work, we will present more discussions and compare these two works in detail in the main text.

It is also argued that the partition of the environment (equivalently, the decomposition of the Hilbert space into tensor products) should be fixed~\cite{QDar1}. In Ref.~\cite{QTP}, a proposition has been proposed that the decomposition of the Hilbert space should be induced by measurable observables. This proposition essentially implies that the decomposition is determined by what measurements people can perform. So it is not suitable to apply in quantum Darwinism (which is aiming at explaining the emergence of classical properties) context.

In this work, we reconsider this ambiguity from a different perspective by studying the entanglement structure of the state of the whole system. We find a uniqueness theorem (named as the extended tridecomposition uniqueness theorem (ETUT)) which applies to a particular subset of all possible partitions of the whole system. Within this subset, the observable of the system leaving redundant imprints in the environment is ``unique". (The precise meaning of this statement will be given in subsection \ref{subsec3-1}.) By ``redundant", we mean that the redundancy is larger or equal to 2. In a special case, when every fraction of the environment is accessible to one local observer for the partitions within the subset that the uniqueness theorem applies, for partitions outside this subset, every fraction of the environment is only accessible to a nonlocal observer (the meaning of which will be explained later), which has no physical significance. This helps us to gain further insight of the emergence of the objectivity problem. In addition, based on the theorem, we can reproduce the relevant results given in Ref.~\cite{Env0}. The benefit of our method is that the results are free from dynamical details or model details. Our results presented here can also be applied to the spectrum broadcast structure framework. In addition, the example elucidated by Eqs. (\ref{Am_ex0}) and (\ref{Am_ex1}) suggests that there is a connection between the present work and another interesting topic, namely, simultaneous measurement on non-commuting observables, which has been investigated in dynamical models~\cite{MeasNon0,MeasNon1,MeasNon2} and experimentally~\cite{MeasNon3,MeasNon4} (see also references therein).

The remaining part of this paper is organized as follows. In section \ref{Sec2}, we set up conventions and introduce the extended tridecomposition uniqueness theorem. In section \ref{Sec3}, based on this theorem, we study the problem whether and on what condition only the pointer observable can leave redundant imprints in the environment. The proof of the theorem is presented in section \ref{Sec4}, and section \ref{Sec5} is devoted to concluding remarks.

\section{The extended tridecomposition uniqueness theorem}\label{Sec2}
\subsection{Definitions and Conventions}\label{subsec2-1}
Consider a macroscopic object bathed in its environment. We will call the object the system of interest or simply the system, but refer to the system$+$environment as the whole system. Both the system of interest and its environment are composed of numerous elementary systems, such as atoms, photons, etc.. So the Hilbert space of the whole system is constructed by the direct product of the Hilbert spaces of all the elementary systems :$H=\bigotimes_i H_i^{ES}$. Although the system of interest is usually described by a density matrix, the whole system can be described by a state vector in the whole Hilbert space. Now let us consider the general situation without making the distinction between the system and the environment. The whole system can be partitioned into subsystems or fractions in different ways, correspondingly, the Hilbert space can be decomposed into subspaces in different ways. For a given whole system, any two possible partitions can be clarified by writing the decompositions of the Hilbert space as
\begin{equation}\label{Pa0}
A:H=H_A^{(1)}\otimes H_A^{(2)}\otimes\dots \otimes H_A^{(n)},
\end{equation}
\begin{equation}\label{Pa1}
B:H=H_B^{(1)}\otimes H_B^{(2)}\otimes\dots \otimes H_B^{(m)},
\end{equation}
and specifying each fractions $H_{A(B)}^{(i)}$ (we use the same notation for a fraction and its corresponding Hilbert subspace). For a given fraction in Partition A (PAT.A), say $H_A^{(i)}$, its elementary systems must be distributed in one or several fractions of Partition B (PAT.B): $H_B^{(j_1)}$, $H_B^{(j_2)}$, $\dots$, $H_B^{(j_l)}$. In other words, elementary systems of $H_A^{(i)}$ belong to (and no more than) $H_B^{(j_1)}$, $H_B^{(j_2)}$, $\dots$, $H_B^{(j_l)}$ when looked in PAT.B. We denote the set of fractions $H_B^{(j_1)}$, $H_B^{(j_2)}$, $\dots$, $H_B^{(j_l)}$ as $\xi_i^{A\rightarrow B}=\{H_B^{(j_1)}, H_B^{(j_2)}, \dots, H_B^{(j_l)}\}$. If we imagine that each elementary system is located on a site in space, then each fraction of a partition defines a region (not necessarily connected) in space, and $\xi_i^{A\rightarrow B}=\{H_B^{(j_1)}, H_B^{(j_2)}, \dots, H_B^{(j_l)}\}$ implies that the region defined by $H_A^{(i)}$ overlaps with the regions defined by $H_B^{(j_1)}, H_B^{(j_2)}, \dots, H_B^{(j_l)}$ and is disjoint from all the other regions of PAT.B, or simply $H_A^{(i)}$ overlaps with $H_B^{(j_1)}, H_B^{(j_2)}, \dots, H_B^{(j_l)}$.

For example, a whole system consisting of six elementary systems can be partitioned in two ways as $A: \{1|2,3,4|5,6\}$ and $B:\{1|2,4|3,5,6\}$, where the numbers represent the elementary systems, and the fractions are separated with the vertical lines. Correspondingly, the whole Hilbert space is decomposed in two ways as
\begin{equation}\label{Pa_ex0}
A:H = H_A^{(1)}\otimes H_A^{(2)}\otimes H_A^{(3)},
\end{equation}
\begin{equation}\label{Pa_ex1}
B:H = H_B^{(1)}\otimes H_B^{(2)}\otimes H_B^{(3)},
\end{equation}
where $H_A^{(2)}=H_2^{ES}\otimes H_3^{ES}\otimes H_4^{ES}$, $H_A^{(3)}=H_5^{ES}\otimes H_6^{ES}$; $H_B^{(2)}=H_2^{ES}\otimes H_4^{ES}$, $H_B^{(3)}=H_3^{ES}\otimes H_5^{ES}\otimes H_6^{ES}$; and $H_A^{(1)}=H_B^{(1)}=H_1^{ES}$. The elementary systems of fraction $H_A^{(2)}$ , i.e., $\{2,3,4\}$ are distributed in $H_B^{(2)}$ and $H_B^{(3)}$ when shown in partition B. So $\xi_2^{A\rightarrow B}$ denotes the set $\{H_B^{(2)},H_B^{(3)}\}$.

{\textit{Definitions:}} We say PAT.B is \textit{comparable} to PAT.A if there exists at least one pair of fractions in PAT.B, say $H_B^{(j_1)}$ and $H_B^{(j_2)}$ ($j_1\neq j_2$), such that the corresponding sets $\xi_{j_1}^{B\rightarrow A}$ and $\xi_{j_2}^{B\rightarrow A}$ satisfy (a) $\xi_{j_1}^{B\rightarrow A} \bigcap \xi_{j_2}^{B\rightarrow A}=\O$ and (b) $\xi_{j_1}^{B\rightarrow A} \bigcup \xi_{j_2}^{B\rightarrow A}\subsetneq \{A\}$, where $\{A\}\equiv \{H_A^{(1)},H_A^{(2)}\dots H_A^{(n)}\}$ denotes the set of all fractions of PAT.A. If PAT.A is also comparable to PAT.B, they are mutually \textit{comparable}. For a given partition, say PAT.A, all the partitions that are mutually comparable to PAT.A including PAT.A itself form a set, which is called the \textit{comparable set} of PAT.A.

\begin{figure}[h]
	\centering	
	\includegraphics[width=0.6\textwidth]{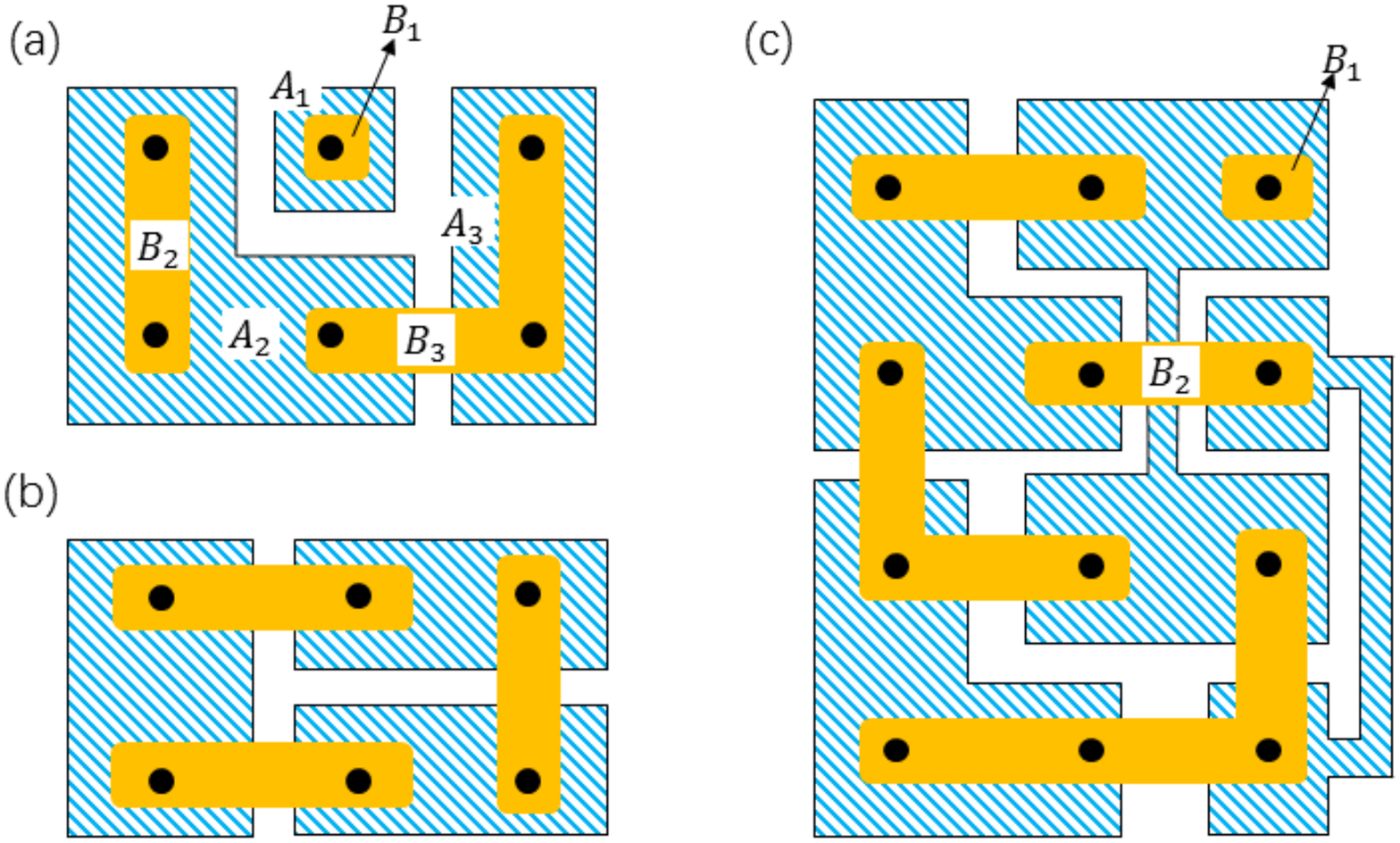}
	\caption{Examples illustrating the idea of comparable sets. Black dots represent elementary systems. A partition is denoted by disjoint regions (each region represents one fraction of this partition) with the same coloring. Specifically, PAT.A is denoted by hashed blue regions and PAT.B is denoted by solid orange regions. (a) The example detailed in the text. PAT.A and PAT.B are mutually comparable. (b) Neither of the partitions are comparable to the other. (c) PAT.B is comparable to PAT.A, which can be checked by looking at $\xi_{1}^{B\rightarrow A}$ and $\xi_{2}^{B\rightarrow A}$, while PAT.A is {\it not} comparable to PAT.B. (Notice that the region denoted by $B_2$ overlaps with two---not three---regions of PAT.A.) }	
	\label{pic1}
\end{figure}

According to the definition, the partitions inside a comparable set must be m-partitions with $m\ge3$. In the example given above, the two partitions are mutually comparable. This example is visualized schematically in Fig.~\ref{pic1} (a). To give an intuitive feeling about the idea of comparable sets, we present two other examples which are described schematically in Fig.~\ref{pic1} (b) and~\ref{pic1} (c). In Fig.~\ref{pic1} (b), neither of the partitions are comparable to the other. In Fig.~\ref{pic1} (c), PAT.B is comparable to PAT.A, while PAT. A is not comparable to PAT. B.

\subsection{The Theorem}\label{subsec2-2}
The physical situation we are interested in is that the whole state of system-environment has become entangled after decoherence. Since entanglement is only meaningful w.r.t. particular partitions of the whole system, a state entangled w.r.t. one partition may be unentangled or entangled in a different manner w.r.t. another partition. The following theorem clarifies some aspects of this problem.

\textit{Definition:} For a partition of the whole system denoted as PAT.A: $H=H_A^{(1)}\otimes H_A^{(2)}\otimes\dots \otimes H_A^{(n)}$ ($n\ge 3$), if a state can be written as $|\Psi\rangle = \sum_{i=1}^{\bar{i}} \alpha_i |\psi_i^{(1)}\rangle|\psi_i^{(2)}\rangle\dots|\psi_i^{(n)}\rangle$, where each set of $|\psi_i^{(j)}\rangle$ is a linearly independent and normalized set in the subspace $H_A^{(j)}$, every $\alpha_i$ is nonzero and $\bar{i}\ge2$, then $|\Psi\rangle$ is called a \textit{semi-GHZ state} w.r.t. PAT.A.

\textit{Theorem:} If a state is a semi-GHZ state w.r.t. PAT.A: $|\Psi\rangle = \sum_{i=1}^{\bar{i}} \alpha_i |\psi_i^{(1)}\rangle|\psi_i^{(2)}\rangle\dots|\psi_i^{(n)}\rangle$, then within the comparable set of PAT.A, the state {\textit{can not be}} a semi-GHZ state w.r.t. a different partition, say PAT.B: $\sum_{i=1}^{\bar{i'}} \kappa_i |\phi_i^{(1)}\rangle|\phi_i^{(2)}\rangle\dots|\phi_i^{(m)}\rangle$, unless $|\psi_i^{(1)}\rangle|\psi_i^{(2)}\rangle\dots|\psi_i^{(n)}\rangle=|\phi_i^{(1)}\rangle|\phi_i^{(2)}\rangle\dots|\phi_i^{(m)}\rangle$ for every $i$ (up to a phase), with a proper ordering of the expanding bases.

This theorem is an extension to the tridecomposition uniqueness theorem which works for a fixed partition~\cite{Uniq}, so we will call this theorem the \textit{extended tridecomposition uniqueness theorem}. The proof of the theorem is given in Section \ref{Sec4}.

To check whether the ETUT applies or not, one needs to check whether two partitions are mutually comparable or not. There exists a sufficient condition for one partition to be comparable to another. Now, for a given partition C, denote the number of elementary systems in one fraction $(j)$ as $L_C^{(j)}$, and the largest(smallest) $L_C^{(j)}$ among all fractions as $L_C^{max}$($L_C^{min}$). One partition, say PAT.B, is comparable to another one, say PAT.A, where PAT.A is an n-partition and PAT.B is an m-partition, if
\begin{equation}\label{sufcon}
m>L_B^{min}(L_A^{max}-1)+1 ~~\mathrm{and}~~n>L_B^{min}+L_B^{max}
\end{equation}
To justify this, consider a fraction in PAT.B with minimum number of elementary systems, and denote it as $H_B^{(1)}$. $H_B^{(1)}$ at most overlaps $L_B^{min}$ fractions in PAT.A (these fractions form $\xi_{1}^{B\rightarrow A}$). For every fraction in the set $\xi_{1}^{B\rightarrow A}$, say $H_A^{(j)}$, it at most overlaps $L_A^{(j)}-1$ fractions in PAT.B other than $H_B^{(1)}$. Then all the fractions in $\xi_{1}^{B\rightarrow A}$ joined overlap at most $L_B^{min}(L_A^{max}-1)+1$ fractions in PAT.B. So if $m>L_B^{min}(L_A^{max}-1)+1$, then there must exist another fraction in PAT.B, say $H_B^{(2)}$, such that it overlaps none of the fractions in $\xi_{1}^{B\rightarrow A}$, i.e., $\xi_{1}^{B\rightarrow A} \bigcap \xi_{2}^{B\rightarrow A}=\O$. Moreover, since $H_B^{(2)}$ consists of at most $L_B^{max}$ elementary systems, if $n>L_B^{min}+L_B^{max}$, then $\xi_{1}^{B\rightarrow A} \bigcup \xi_{2}^{B\rightarrow A}\subsetneq \{A\}$ must holds. So Eq.~(\ref{sufcon}) implies that PAT.B is comparable to PAT.A.

\section{physical implications of the theorem}\label{Sec3}
\subsection{Perfect Correlation}\label{subsec3-1}
Now consider a system of interest has decohered by its environment. The pointer bases of the system are represented by $|\psi_i^{(S)}\rangle$ ($i=1,2,\dots,d_S$), where $d_S$ is the dimensionality of the system. Suppose a perfect correlation among the system and the fractions of the environment has established, and the state has a GHZ-like form w.r.t. a particular partition:
\begin{equation}\label{De0}
|\Psi\rangle = \sum_{i=1}^{\bar{i}} \alpha_i |\psi_i^{(S)}\rangle|\psi_i^{(1)}\rangle\dots|\psi_i^{(n_e)}\rangle,
\end{equation}
where $|\psi_i^{(j)}\rangle$'s form a set of orthonormal bases of the Hilbert subspace of the environmental fraction $j$. In Eq. (\ref{De0}) every $\alpha_i$ is nonzero and $\bar{i}$ is no greater than the minimum dimensionality among the system and every environmental fraction. Denote this partition as PAT.A which is clarified by $H=H_A^{(S)}\otimes H_A^{(1)}\otimes\dots \otimes H_A^{(n_e)}$, where $H_A^{(S)}$ represents the Hilbert space of the system and $H_A^{(j)}$ represents the Hilbert space of the $j$th environmental fraction. (If we consider the density matrix $|\Psi\rangle\langle\Psi|$ and trace out one of the fractions of the environment, we obtain a so-called spectrum broadcast structure~\cite{SBS0}.)

The structure of the state shows that the information of the pointer observable has proliferated in the environment, which can be seen by evaluating the classical mutual information between the system and each fraction of the environment. The classical mutual information of an observable $\hat{S}$ in the Hilbert space of the system and an observable $\hat{F}$ in the Hilbert space of an environmental fraction is defined as~\cite{Env0}
\begin{equation}
I(\hat{S}:\hat{F})=H(\hat{S})-H(\hat{S}|\hat{F}),
\end{equation}
where $H(\hat{S})\equiv -\sum_i p(S_{i})\log_2 p(S_{i})$ is the Shannon entropy, and $p(S_{i})$ is the probability associated with the measurement outcome $S_{i}$. $H(\hat{S}|\hat{F})=-\sum_{ij}p(F_{j})p(S_{i}|F_j)\log_2 p(S_{i}|F_j)$, where $p(S_{i}|F_j)$ is the conditional probability of $S_i$ given $F_j$. The maximum $I(\hat{S}:\hat{F})=H(\hat{S})$ occurs when $H(\hat{S}|\hat{F})=0$, which means we can gain the complete information about $\hat{S}$ by measuring $\hat{F}$. Now introduce the pointer observable $\hat{\Lambda}=\sum_i \lambda_i |\psi_i^{(S)}\rangle\langle\psi_i^{(S)}|$, where $\lambda_i$'s are all real and nonequal, and an observable of the environmental fraction $j$: $\hat{\Omega}=\sum_i \omega_i |\psi_i^{(j)}\rangle\langle\psi_i^{(j)}|$, where $\omega_i$'s are all real and nonequal. (We only consider the non-degenerate case because $\hat{\Lambda}$ and $\hat{\Omega}$ could represent a complete set of commuting observables in their corresponding subspaces. There is a one-to-one correspondence between $\lambda_i(\omega_i)$ and the set of eigenvalues of the commuting observables.) From Eq. (\ref{De0}), it is easy to see that $H(\hat{\Lambda}|\hat{\Omega})=0$, so the pointer observable leaves a perfect imprint in the fraction $j$, which is also true for any other environmental fraction in PAT.A. On the other hand, if $H(\hat{\Lambda}|\hat{F}^{(j)})=0$ and $H(\hat{F}^{(j)}|\hat{\Lambda})=0$ for some nondegenerate observable $\hat{F}^{(j)}$ acting on the $j$th fraction and for every $j$, then the information of $\hat{\Lambda}$ is recorded nondegenerately, and the state $|\Psi\rangle$ must be able to be expressed in a GHZ-like form as Eq. (\ref{De0}). We are restricted to this situation in this paper. So, every GHZ-like expression of a given state corresponds to an information (perfect) multiplication of a particular observable.

Given Eq. (\ref{De0}), for PAT.A, there is no alternative expansion of $|\Psi\rangle$ having a GHZ-like structure according to the tridecomposition uniqueness theorem~\cite{Uniq}, so only the pointer observable $\hat{\Lambda}$ has redundant imprints in the environment w.r.t. PAT.A. But if different partitions of environment are possible, then the state may have a different GHZ-like expression (as shown by the example displayed in Introduction), and a different observable of the system may leave multiple imprints in the environment as well. Regarding this problem, the extended tridecomposition uniqueness theorem tells that, as long as the partitions are restricted within the comparable set of PAT.A, there is no alternative expansion of $|\Psi\rangle$ having a GHZ-like structure. Thus, within the comparable set of PAT.A, only the pointer observable $\hat{\Lambda}$ has redundant imprints in the environment. More precisely, the observable leaving redundant and nondegenerately recorded imprints in the environment is ``unique", where ``unique" means, only those observables whose eigenstates include all the $|\psi_i^{(S)}\rangle$'s appearing in $|\Psi\rangle$ may leave redundant imprints in the environment.

Then, how about the partitions outside the comparable set of PAT.A? We argue that, if for PAT.A, every fraction of the environment is accessible to a local observer, then for any partition (with one subsystem being the system of interest) outside the comparable set of PAT.A, every fraction of the environment is only accessible to a nonlocal observer, where a nonlocal observer is an observer that is able to gain the information about the whole environment. To explain this, we first show what mutual comparability implies in the present case. Consider another partition denoted as PAT.B: $H=H_B^{(S)}\otimes H_B^{(1)}\otimes\dots \otimes H_B^{(m_e)}$. Since we are considering the partitions with the system of interest fixed, we have $H_A^{(S)}=H_B^{(S)}$ and $\xi_S^{B\rightarrow A}=\{H_A^{(S)}\}$ and vice versa. And it is obvious that $\xi_j^{B\rightarrow A}\bigcap\xi_S^{B\rightarrow A}=\O$ for any fraction $j$ of the environment w.r.t. PAT.B. So, if $\xi_j^{B\rightarrow A} \subsetneq \{H_A^{(1)},H_A^{(2)},\dots,H_A^{(n_e)}\}$ for at least one $j$, then PAT.B is comparable to PAT.A; on the other hand, if $\xi_j^{B\rightarrow A}= \{H_A^{(1)},H_A^{(2)},\dots,H_A^{(n_e)}\}$ for every $j$ then PAT.B is \textit{not} comparable to PAT.A. In addition, one can prove that if $\xi_j^{B\rightarrow A}= \{H_A^{(1)},H_A^{(2)},\dots,H_A^{(n_e)}\}$ for every $j$, then $\xi_j^{A\rightarrow B}= \{H_B^{(1)},H_B^{(2)},\dots,H_B^{(m_e)}\}$ for every $j$. So, PAT.B is not comparable to PAT.A if and only if $\xi_j^{B\rightarrow A}= \{H_A^{(1)},H_A^{(2)},\dots,H_A^{(n_e)}\}$ for every $j$.

The physical meaning of the condition $\xi_j^{B\rightarrow A}= \{H_A^{(1)},H_A^{(2)},\dots,H_A^{(n_e)}\}$ for every $j$ is that the elementary systems in every fraction of the environment in PAT.B must be distributed in all the fractions of the environment in PAT.A. When the environment is divided into fractions each accessible to a local observer, an observer that is able to receive information from all these fractions should be a nonlocal observer, because in this situation there is no reason to forbid the observer to gain information of the whole environment. So, if every environmental fraction of PAT.A is accessible to a local observer, then every environmental fraction of PAT.B can only be ``observed" by a nonlocal observer. To get an intuitive picture, consider a macroscopic object decohered by its environmental photons. The photons spread into a wide open space so no local observer can receive all of them, but a small fraction of them which occupies a location with finite extension could be received by a local observer. Dividing the whole environment into such fractions, then an observer who receives photons from all the fractions must occupy the whole space that the environment spreads into, so this observer is able to receive the whole environmental photons and is a nonlocal observer.
Recalling that for decoherence theory and quantum Darwinism to work, nonlocal observers should be excluded, because for a nonlocal observer ``observing" the whole state, decoherence had never happened. Since in the present case, every environmental fraction of PAT.B is only accessible to a nonlocal observer, this partition has no physical significance. Although we use the word ``observer" in our arguments, it does not necessarily refer to a live being. A local observer is a physical system with finite extension, which can only communicate with part of the environment; on the contrary, a nonlocal observer is the physical system that can communicate with the whole system.

The situation discussed before is common in the real world, but in principle, there could be more sophisticated situations such as when not all of the environmental fractions in PAT.A are accessible to local observers. In such situations, we still need the requirement already indicated in Ref.~\cite{Env0} that the redundancy should be sufficiently large to guarantee no occurrence of ambiguity. To give a quantitative criterion, introduce the redundancy of an operator $\hat{X}$ (acting on the Hilbert space of the system of interest): $R_{\delta}(\hat{X})$, defined as the number of the disjoint fractions of the environment containing all but a fraction $\delta$ of the information about $\hat{X}$ present in the entire environment. In the perfect correlation case, we have $\delta=0$ and the redundancy of the pointer observable is $R_0(\hat{\Lambda})$. Implied by Eq. (\ref{De0}), the redundancy $R_0(\hat{\Lambda})$ is equal or greater than $n_e$. When $R_0(\hat{\Lambda})$ is greater than $n_e$, we could fine-grain the partition of the environment and eventually, we could write the state in a form such that each fraction can no longer be finer-grained while still contain the whole information of $\hat{\Lambda}$. Suppose Eq. (\ref{De0}) is the finest-grained in this sense, we have $R_0(\hat{\Lambda})=n_e$. If there is another partition, say PAT.B, with respect to which the state $|\Psi\rangle$ can have a (finest-grained) GHZ-like form, then there is another observable $\hat{\Lambda}'$ whose redundancy is $R_0(\hat{\Lambda})=m_e$ where $m_e$ is the number of environmental fractions. Suppose the total number of elementary systems of the environment is $N_e$, in order for these two partitions to satisfy the requirements to apply the ETUT, we only need $n_e>L_{Bmin}$ or $m_e>L_{Amin}$, where $L_{Amin}(L_{Bmin})$ represents the minimum of the numbers of elementary systems in each fraction of PAT.A(B). Noticing that $L_{Amin}\le \frac{N_e}{n_e}$ and $L_{Bmin}\le\frac{N_e}{m_e}$, so if $n_e m_e> N_e$ or equivalently $R_0(\hat{\Lambda})R_0(\hat{\Lambda}')> N_e$, then the requirements would be satisfied and the ETUT applies, which means that the two GHZ-like expressions of $|\Psi\rangle$ w.r.t. PAT.A and PAT.B must be identical and $[\hat{\Lambda},\hat{\Lambda}']=0$. It follows immediately that if we only admit redundancies satisfying $R_0>\sqrt{N_e}$, then the observable leaving redundant imprints in the environment is ``unique". This is exactly the result obtained in Ref. \cite{Env0}. We obtain it using a different method.

\subsection{Imperfect Correlation}
In most realistic situations, decoherence processes only induce a fast damping of the coherent phases, and the perfect decoherence never occur in finite time. So it is important to study the case when the correlation among the system and the environmental fragments is imperfect. In this case, we can express the state as
\begin{equation}\label{De_im0}
|\Psi\rangle = \sum_{i=1}^{\bar{i}} \alpha_i |\psi_i^{(S)}\rangle|\psi_i^{(1)}\rangle\dots|\psi_i^{(n_e)}\rangle+\sum_{\{i\}} \beta_{\{i\}} |\psi_{is}^{(S)}\rangle|\psi_{i1}^{(1)}\rangle\dots|\psi_{in_e}^{(n_e)}\rangle,
\end{equation}
where $\{i\}$ denotes $\{is,i1,\dots,in_e\}$ excluding the cases when $is=i1=\cdots=in_e$. Such an expansion for any state is always possible because all the states $|\psi_{is}^{(S)}\rangle|\psi_{i1}^{(1)}\rangle\dots|\psi_{in_e}^{(n_e)}\rangle$ form a complete set of orthonormal bases in the whole Hilbert space. We call each of the bases a product basis w.r.t. PAT.A. Only when $\sum_{\{i\}} |\beta_{\{i\}}|^2$ is finite and $\sum_{\{i\}} |\beta_{\{i\}}|^2 \ll \sum_i |\alpha_i|^2 $, the state is said to have an approximate GHZ-like form w.r.t. PAT.A, and the observable $\hat{\Lambda}$ leaves redundant but imperfect imprints with redundancy $R_\delta(\hat{\Lambda})\ge n_e$ in the environment.

Before we draw any conclusion, let's consider the following situation first. Suppose a state $|\Psi_0\rangle$ has a perfect correlation among the system and the environmental fractions w.r.t. PAT.A. This state can also be expanded in another complete set of product bases w.r.t. PAT.B as
\begin{eqnarray}\label{De_im1}
|\Psi_0\rangle &=& \sum_{i=1}^{\bar{i}} \alpha_i |\psi_i^{(S)}\rangle|\psi_i^{(1)}\rangle\dots|\psi_i^{(n_e)}\rangle \notag\\
&=&\sum_{i=1}^{\bar{i}'} \alpha'_i |\phi_i^{(S)}\rangle|\phi_i^{(1)}\rangle\dots|\phi_i^{(m_e)}\rangle+\sum_{\{i\}} \beta'_{\{i\}} |\phi_{is}^{(S)}\rangle|\phi_{i1}^{(1)}\rangle\dots|\phi_{im_e}^{(m_e)}\rangle,
\end{eqnarray}
where for each $x$, $|\phi_{ix}^{(x)}\rangle$'s form a complete set of orthonormal bases in the Hilbert space of the subsystem $x$. For PAT.B, there are infinitely many complete sets of product bases. For the same bases, we serve different ordering as forming different sets. (For instance, for a bipartite whole system, if one of its complete sets of product bases is $\{|\phi_{1}^{(1)}\rangle|\phi_{1}^{(2)}\rangle, |\phi_{2}^{(1)}\rangle|\phi_{2}^{(2)}\rangle, |\phi_{1}^{(1)}\rangle|\phi_{2}^{(2)}\rangle, |\phi_{2}^{(1)}\rangle|\phi_{1}^{(2)}\rangle\}$, then after reordering $|\phi_{i}^{(2)}\rangle$, i.e. $|\phi_{1\leftrightarrow2}^{(2)}\rangle$, the new set is considered as a different one.) When varying the product bases, the coefficients $\alpha'_i$ and $\beta'_{\{i\}}$ vary accordingly. In a complete set, the $k$th product basis can be described by its wave function, i.e., $C^k(i)$, in a reference representation. Then $\delta_2\equiv \sum_{\{i\}}|\beta'_{\{i\}}|^2$ is an analytic function of $C^k(i)$'s. For PAT.B within the comparable set of PAT.A, the ETUT tells that $\delta_2=0$ only when the set of $|\psi_i^{(S)}\rangle|\psi_i^{(1)}\rangle\dots|\psi_i^{(n_e)}\rangle$ and the set of $|\phi_i^{(S)}\rangle|\phi_i^{(1)}\rangle\dots|\phi_i^{(m_e)}\rangle$ coincide (up to a phase). So for sufficiently small $\delta_2$, the set of $|\phi_i^{(S)}\rangle|\phi_i^{(1)}\rangle\dots|\phi_i^{(m_e)}\rangle$ must be \textit{approximately} coincide with the set of $|\psi_i^{(S)}\rangle|\psi_i^{(1)}\rangle\dots|\psi_i^{(n_e)}\rangle$. To be precise, writing
\begin{equation}\label{eps}
\langle\psi_i^{(n_e)}|\dots\langle\psi_i^{(1)}|\langle\psi_i^{(S)}|\phi_j^{(S)}\rangle|\phi_j^{(1)}\rangle\dots|\phi_j^{(m_e)}\rangle=e^{i\theta_j}(\delta_{ij}+\epsilon_{ij}),
\end{equation}
then for sufficiently small $\delta_2$, we must have $|\epsilon_{ij}|\ll 1$ for every legal $i,j$ under a proper ordering.

Now, go back to the general situation Eq. (\ref{De_im0}) when the state has an approximate GHZ-like form. Expanding it in another complete set of product bases w.r.t. PAT.B:
\begin{eqnarray}\label{De_im2}
|\Psi\rangle &=& \sum_{i=1}^{\bar{i}} \alpha_i |\psi_i^{(S)}\rangle|\psi_i^{(1)}\rangle\dots|\psi_i^{(n_e)}\rangle+\sum_{\{i\}} \beta_{\{i\}} |\psi_{is}^{(S)}\rangle|\psi_{i1}^{(1)}\rangle\dots|\psi_{in_e}^{(n_e)}\rangle \notag\\
&=&\sum_{i=1}^{\bar{i}'} \alpha'_i |\phi_i^{(S)}\rangle|\phi_i^{(1)}\rangle\dots|\phi_i^{(m_e)}\rangle+\sum_{\{i\}} \beta'_{\{i\}} |\phi_{is}^{(S)}\rangle|\phi_{i1}^{(1)}\rangle\dots|\phi_{im_e}^{(m_e)}\rangle.
\end{eqnarray}
Moving the term $\sum_{\{i\}} \beta_{\{i\}} |\psi_{is}^{(S)}\rangle|\psi_{i1}^{(1)}\rangle\dots|\psi_{in_e}^{(n_e)}\rangle$ to the right hand side of the second equality sign, and expanding it in the $|\phi_{is}^{(S)}\rangle|\phi_{i1}^{(1)}\rangle\dots|\phi_{im_e}^{(m_e)}\rangle$ bases, we can make the same argument as we did for Eq. (\ref{De_im1}). It turns out that for sufficiently small $\sum_{\{i\}}|\beta_{\{i\}}|^2+\sum_{\{i\}}|\beta'_{\{i\}}|^2$, $|\psi_i^{(S)}\rangle|\psi_i^{(1)}\rangle\dots|\psi_i^{(n_e)}\rangle$ and $|\phi_i^{(S)}\rangle|\phi_i^{(1)}\rangle\dots|\phi_i^{(m_e)}\rangle$ approximately coincide, i.e., $|\epsilon_{ij}|\ll 1$ for every legal $i,j$, under a proper ordering. Writing $\langle\psi_i^{(S)}|\phi_j^{(S)}\rangle=e^{i\theta^s_j}(\delta_{ij}+\epsilon^s_{ij})$, $\langle\psi_i^{(n_e)}|\dots\langle\psi_i^{(1)}|\phi_j^{(1)}\rangle\dots|\phi_j^{(m_e)}\rangle=e^{i\theta^{\bar{s}}_j}(\delta_{ij}+\epsilon^{\bar{s}}_{ij})$, we obtain $e^{i\theta_j}(\delta_{ij}+\epsilon_{ij})=e^{i\theta^s_j}(\delta_{ij}+\epsilon^s_{ij})e^{i\theta^{\bar{s}}_j}(\delta_{ij}+\epsilon^{\bar{s}}_{ij})
=e^{i(\theta^s_j+\theta^{\bar{s}}_j)}(\delta_{ij}
+\delta_{ij}\epsilon^s_{ij}+\delta_{ij}\epsilon^{\bar{s}}_{ij}+\epsilon^s_{ij}\epsilon^{\bar{s}}_{ij})$. Thus, $\epsilon^s_{ii}$ is (at most) of the same order as $\epsilon_{ii}$, i.e., $|\epsilon^s_{ii}|\ll 1$. So, (within the comparable set of PAT.A), the states having redundant imperfect imprints in the environment are approximately the pointer states.

\subsection{Comparisons with Ref.~\cite{Loc}}
In Ref.~\cite{Loc}, a similar issue was addressed. In particular, the author proved a theorem (termed ``main result" in that paper), which identifies a particular decomposition of the state with the feature quite similar to the GHZ-type entanglement we concentrate on here. So, we give close comparisons between the main results of Ref.~\cite{Loc} and the present work. We assume that the readers are already familiar with the work of Ref.~\cite{Loc}. To make connections of these two works, we indicate that the term ``region" (where local observables are defined on) used in Ref.~\cite{Loc} corresponds to ``fraction" in our terminology. The set of disjoint regions on which a redundantly recorded observable is defined in Ref.~\cite{Loc} corresponds to a particular partition in this paper. (A difference is that, in Ref.~\cite{Loc}, the disjoint regions may not exhaust the whole space, while in this work, a partition always exhausts the whole space.) A key concept introduced in Ref.~\cite{Loc} is the so-called ``pair-cover". Translating this term into our terminology, we find that a redundantly recorded observable $\hat{\Omega}_A$ (corresponding to PAT.A) pair-covers another one $\hat{\Omega}_B$ (corresponding to PAT.B) means that there is at least one pair of fractions of PAT.A, say $j_1,j_2$, such that $\xi_{j_1}^{A\rightarrow B} \bigcup \xi_{j_2}^{A\rightarrow B}= \{B\}$. Equivalently, that $\hat{\Omega}_A$ does not pair-cover $\hat{\Omega}_B$ corresponds to $\xi_{j_1}^{A\rightarrow B} \bigcup \xi_{j_2}^{A\rightarrow B}\subsetneq \{B\}$ for {\it every pair} of fractions in PAT.A. The main result of Ref.~\cite{Loc} states that given a collection of recorded observables for the state of the whole system $|\Psi\rangle$, if none of the recorded observables pair-covers another, then they define a joint branch decomposition of simultaneous eigenstates of all records. Now we can discuss the main differences between the ``main result" of Ref.~\cite{Loc} and the extended tridecomposition uniqueness theorem of this work as follows.
\begin{description}
\item[(a)] The condition assumed in the ``main result" of Ref.~\cite{Loc} is different from the condition assumed in the ETUT. To be specific, that two redundantly recorded observables do not pair-cover each other means $\xi_{j_1}^{A\rightarrow B} \bigcup \xi_{j_2}^{A\rightarrow B}\subsetneq \{B\}$ for {\it every pair} of fractions in PAT.A and vise versa. On the other hand, in the ETUT, it is assumed $\xi_{j_1}^{A\rightarrow B} \bigcup \xi_{j_2}^{A\rightarrow B}\subsetneq \{B\}$ for {\it at least one pair} of fractions in PAT.A accompanied with the condition $\xi_{j_1}^{A\rightarrow B} \bigcap \xi_{j_2}^{A\rightarrow B}=\O$, and vise versa. Although there are many cases where the two conditions both hold true, in general, neither the conditions implies the other. In the example presented in Fig.~\ref{pic1} (a), the two partitions are mutually comparable, so the ETUT applies, while the corresponding redundantly recorded observables pair-cover each other, so the ``main result" of Ref.~\cite{Loc} does not follow. Examples that two redundantly recorded observables do not pair-cover each other while the corresponding partitions are not mutually comparable also exist.\footnote{Here is an example. Suppose PAT.A has six fractions (regions), PAT.B has seven fractions (regions). The overlaps between the regions of the two partitions are illustrated by $\xi_{1}^{A\rightarrow B}=\{H_B^{(1)},H_B^{(2)},H_B^{(3)}\}$, $\xi_{2}^{A\rightarrow B}=\{H_B^{(1)},H_B^{(4)},H_B^{(5)}\}$, $\xi_{3}^{A\rightarrow B}=\{H_B^{(1)},H_B^{(6)},H_B^{(7)}\}$, $\xi_{4}^{A\rightarrow B}=\{H_B^{(2)},H_B^{(4)},H_B^{(5)},H_B^{(6)}\}$, $\xi_{5}^{A\rightarrow B}=\{H_B^{(2)},H_B^{(3)},H_B^{(5)},H_B^{(7)}\}$, $\xi_{6}^{A\rightarrow B}=\{H_B^{(3)},H_B^{(4)},H_B^{(6)},H_B^{(7)}\}$ and $\xi_{1}^{B\rightarrow A}=\{H_A^{(1)},H_A^{(2)},H_A^{(3)}\}$, $\xi_{2}^{B\rightarrow A}=\{H_A^{(1)},H_A^{(4)},H_A^{(5)}\}$, $\xi_{3}^{B\rightarrow A}=\{H_A^{(1)},H_A^{(5)},H_A^{(6)}\}$, $\xi_{4}^{B\rightarrow A}=\{H_A^{(2)},H_A^{(4)},H_A^{(6)}\}$, $\xi_{5}^{B\rightarrow A}=\{H_A^{(2)},H_A^{(4)},H_A^{(5)}\}$, $\xi_{6}^{B\rightarrow A}=\{H_A^{(3)},H_A^{(4)},H_A^{(6)}\}$, $\xi_{7}^{B\rightarrow A}=\{H_A^{(3)},H_A^{(5)},H_A^{(6)}\}$. Then the redundant recorded observables associated to PAT.A and PAT.B non-pair-cover each other, while PAT.A is not comparable to PAT.B and vice versa. }

\item[(b)] The ETUT concerns the decomposition of the state $|\Psi\rangle$ into linearly independent components (see subsection \ref{subsec2-2}), while the result proved in Ref.~\cite{Loc} only concerns orthogonal decompositions of the state.

\item[(c)] Restricting to the orthogonal decompositions of $|\Psi\rangle$, the components (or the expanding bases) of the decomposition defined in the ETUT and in the ``main result" of Ref.~\cite{Loc} are still different in general, which can be illustrated by the following example. Suppose the whole system consists of a number of qubits. Consider two partitions: $A:H = \bigotimes_i H_A^{(i)}$ and $B:H = \bigotimes_i H_B^{(i)}$, where $H_A^{(1)}$ consists of qubits 1 and 2, $H_A^{(2)}$ consists of qubits 3 and 4, $H_B^{(1)}$ consists of qubits 1 and 4, $H_B^{(2)}$ consists of qubits 2 and 3, and for the remaining fractions $H_A^{(3)}=H_B^{(3)}, H_A^{(4)}=H_B^{(4)}, \cdots$. Now define local observables $$\hat{\Omega}_{X}=\omega_{X}(|00\rangle\langle00|+|11\rangle\langle11|)_{X}
    +\omega_{X}'(|01\rangle\langle01|+|10\rangle\langle10|)_{X},$$
    where $X$ stands for $A_1$, $A_2$, $B_1$ or $B_2$ and indicates on which fraction (region) the observable is defined.
    Suppose the state $|\Psi\rangle$ is given as \begin{eqnarray}|\Psi\rangle &=& \alpha(|00\rangle_{A_1}|00\rangle_{A_2}+|11\rangle_{A_1}|11\rangle_{A_2})|A_3\rangle|A_4\rangle\cdots
    +\beta(|01\rangle_{A_1}|01\rangle_{A_2}+|10\rangle_{A_1}|10\rangle_{A_2})|A_3'\rangle|A_4'\rangle\cdots \notag\\
    &=&\alpha(|00\rangle_{B_1}|00\rangle_{B_2}+|11\rangle_{B_1}|11\rangle_{B_2})|A_3\rangle|A_4\rangle\cdots
    +\beta(|01\rangle_{B_1}|10\rangle_{B_2}+|10\rangle_{B_1}|01\rangle_{B_2})|A_3'\rangle|A_4'\rangle\cdots\label{Com_dec}\end{eqnarray}
    where $|A_j\rangle$ and $|A_j'\rangle$ are two eigenstates of an observable $\hat{\Omega}_j$ defined on $H_A^{(j)}=H_B^{(j)}$ with distinct eigenvalues for $j\ge 3$. One can check that two sets of observables $\{\hat{\Omega}_{A_1},\hat{\Omega}_{A_2},\hat{\Omega}_{j} (j\ge 3)\}$ and $\{\hat{\Omega}_{B_1},\hat{\Omega}_{B_2},\hat{\Omega}_{j} (j\ge 3)\}$ associated with PAT.A and PAT.B respectively do not pair-cover each other (when the number of total fractions or regions are greater or equal to 4) and are compatible (in the terminology of Ref.~\cite{Loc}) on $|\Psi\rangle$, so they define a decomposition of $|\Psi\rangle$ as shown in Eq.~(\ref{Com_dec}). On the other hand, $|\Psi\rangle$ is not a GHZ-like state w.r.t either PAT.A or PAT.B. (It is a GHZ-like state w.r.t. the partition when $H_{A}^{(1)}$ and $H_{A}^{(2)}$ are united.) This example illustrates that the ETUT concerns the (semi-)GHZ type entanglement of a given state while Ref.~\cite{Loc} concerns the decomposition identified by a set of compatible observables, although there are many situations where the two decompositions coincide. Obviously, the degeneracy of the observables $\hat{\Omega}_X$ plays an important role in this example. While degeneracies of observables are quite general, the ETUT is sufficient to investigate the physical situation we are interested in here. (If we identify the qubits in $A_1$ as the ``system of interest" and all the others as the environment, then the system has decohered by the environment but the number of fractions in PAT.A does not reflect the redundancy of the pointer observable.)
\end{description}
The above discussions show that, neither the assumptions (a) nor the decompositions (b), (c) considered in Ref.~\cite{Loc} and in the ETUT are the same in general.

Regarding the results given in subsection~\ref{subsec3-1}, we make further clarifications as follows. In this work, we apply the ETUT in the context of quantum Darwinism, where one of the fraction is fixed to be the ``system of interest", and the information about the pointer observable has redundantly recorded in the environment, which means a partition (say PAT.A) has emerged with each fraction recording the information of the pointer observable. If every record in the environment is accessible to a local observer, then for the partitions non-comparable to PAT.A, each region must overlaps with all the regions (in the environment) w.r.t. PAT.A, which we call ``accessible to nonlocal observers". So restricting to the comparable set of PAT.A amounts to excluding the situations where every record (of some other observable) is stored ``nonlocally". And within the comparable set of PAT.A, the ETUT implies, loosely speaking, only the pointer observable leaves redundant imprints. This conclusion is less restricted compared to the corollary given in Ref.~\cite{Loc}, which requires each record to be local observable {\it with a fixed extension} and {\it pairwise separated} by a fixed distance. However, our conclusion holds true only when the distinction of local observer and nonlocal observer can be physically justified, while the corollary of Ref. [30] is rigorous.

To sum up, Ref.~\cite{Loc} and the present work revealed related yet different aspects of the problem about seeking a unique decomposition of a state associated with redundant recorded observables.

\section{Proof of the theorem}\label{Sec4}
In this section, we prove the extended tridecomposition uniqueness theorem. We will prove a lemma first, then the theorem is proved based on the lemma.

\textit{Lemma.} Consider a Hilbert space $H$ and its two decompositions PAT.A having $n$ subspaces and PAT.B having $m$ subspaces (see Eqs. (\ref{Pa0},~\ref{Pa1})). A factorized state $|\Psi\rangle=|\psi^{(1)}\rangle|\psi^{(2)}\rangle\dots|\psi^{(n)}\rangle$ w.r.t. PAT.A (denoted as Form A) can \textit{not} be written as $|\Psi\rangle=\sum_{i=1}^{\bar{i}'} \kappa_i |\phi_i^{(1)}\rangle|\phi_i^{(2)}\rangle\dots|\phi_i^{(m)}\rangle$ ($\kappa_i\ne 0$) w.r.t. PAT.B (denoted as Form B) where $\bar{i}'\geq2$, if the following two conditions are satisfied: (a) There exists at least one pair of fractions of PAT.B, say $H_B^{(j_1)}$ and $H_B^{(j_2)}$ ($j_1\neq j_2$), such that the corresponding sets $\xi_{j_1}^{B\rightarrow A}$ and $\xi_{j_2}^{B\rightarrow A}$ satisfy $\xi_{j_1}^{B\rightarrow A} \bigcap \xi_{j_2}^{B\rightarrow A}=\O$. (b) The states $|\phi_i^{(j_1)}\rangle$ ($i=1,2$,$\cdots$) are independent, and the states $|\phi_i^{(j_2)}\rangle$ ($i=1,2$,$\cdots$) are non-collinear, or vice versa.

This lemma is proved by contradiction. Given the assumptions of the lemma, Form A and Form B of $|\Psi\rangle$ imply contradicting entanglement structures between the two parts of a particular bipartition of the state, which can be revealed by evaluating the corresponding entanglement entropy.

\textit{Proof.} Without losing generality, we take $j_1=1$ and $j_2=2$, and $\xi_{1}^{B\rightarrow A}$ ($\xi_{2}^{B\rightarrow A}$) as the set of subsystems $\{H_A^{(1)}, H_A^{(2)}, \dots, H_A^{(l)}\}$ ($\{H_A^{(l+1)}, H_A^{(l+2)}, \dots, H_A^{(l+l')}\}$) of PAT.A. Now, consider the bipartition: $H_B^{(2)}\otimes\overline{H_B^{(2)}}$, where $H_B^{(2)}$ is the subspace associated with the fraction 2 w.r.t. PAT.B, and $\overline{H_B^{(2)}}$ represents the complementary subspace of $H_B^{(2)}$. We will show that the entanglement entropy w.r.t. this bipartition evaluated using Form B alone contradicts that evaluated using Form A and Form B together, which proves the lemma.

First, using Form B: $|\Psi\rangle=\sum_{i=1}^{\bar{i}'} \kappa_i |\phi_i^{(1)}\rangle|\phi_i^{(2)}\rangle\dots|\phi_i^{(m)}\rangle$, one finds
\begin{equation}\label{EEB}
S(\mathrm{Tr}^{(2)}|\Psi\rangle\langle\Psi|)>0,
\end{equation}
where $\mathrm{Tr}^{(2)}$ denotes the partial trace over the subspace $H_B^{(2)}$, and $S(\rho)\equiv -\mathrm{Tr}\rho\log_2\rho$ is the von Neumann entropy of $\rho$. In the present case, $S(\mathrm{Tr}^{(2)}|\Psi\rangle\langle\Psi|)$ is just the entanglement entropy of $|\Psi\rangle$ w.r.t. the bipartition $H_B^{(2)}\otimes\overline{H_B^{(2)}}$.

Eq.~(\ref{EEB}) follows because given the expression $|\Psi\rangle=\sum_{i=1}^{\bar{i}'} \kappa_i |\phi_i^{(1)}\rangle|\phi_i^{(2)}\rangle\dots|\phi_i^{(m)}\rangle$ and the conditions required in the Lemma, $|\Psi\rangle$ cannot be a factorized state w.r.t. the bipartition $H_B^{(2)}\otimes\overline{H_B^{(2)}}$~\cite{Uniq}. So when written into its Schmidt decomposition form $|\Psi\rangle=\sum_i\lambda_i |A_i\rangle|B_i\rangle$, $|\Psi\rangle$ must have the Schmidt rank larger than 1 with $\lambda_i<1$, so we must have $S(\mathrm{Tr}^{(2)}|\Psi\rangle\langle\Psi|)=-\sum_i\lambda_i^2 \log_2\lambda_i^2 >0$.

Second, using Form A, and noticing that $\xi_{2}^{B\rightarrow A}=\{H_A^{(l+1)}, H_A^{(l+2)}, \dots H_A^{(l+l')}\}$, we have
\begin{equation}
\mathrm{Tr}^{(2)}|\Psi\rangle\langle\Psi|=|\psi^{(1)}\rangle\dots|\psi^{(l)}\rangle\langle\psi^{(l)}|\dots\langle\psi^{(1)}|
\otimes\mathrm{Tr}^{(2)}|\psi^{(l+1)}\rangle\dots|\psi^{(n)}\rangle\langle\psi^{(n)}|\dots\langle\psi^{(l+1)}|.
\end{equation}
Thus,
\begin{equation}\label{Spsi1}
S(\mathrm{Tr}^{(2)}|\Psi\rangle\langle\Psi|)=S(|\psi^{(1)}\rangle\dots|\psi^{(l)}\rangle\langle\psi^{(l)}|\dots\langle\psi^{(1)}|)+S(\rho_2)=S(\rho_2),
\end{equation}
where
\begin{equation}
\rho_2=\mathrm{Tr}^{(2)}|\psi^{(l+1)}\rangle\dots|\psi^{(n)}\rangle\langle\psi^{(n)}|\dots\langle\psi^{(l+1)}|,
\end{equation}
and the relation $S(\rho_1\otimes\rho_2)=S(\rho_1)+S(\rho_2)$ has been used.

To evaluate $S(\rho_2)$, we take advantage of both Form B and Form A of $|\Psi\rangle$.
Suppose the set of states $|\phi_i^{(1)}\rangle$ ($i=1,2$,$\cdots$) is an independent set in subspace $H_B^{(1)}$. One can always find a state orthogonal to all states but $|\phi_1^{(1)}\rangle$ in this subspace. We denote this state as $|\phi_{\perp\bar{1}}\rangle$, and define the projector $\hat{P}_{1}\equiv|\phi_{\perp\bar{1}}\rangle\langle\phi_{\perp\bar{1}}|$. Given Form B of $|\Psi\rangle$, we project out a single component from $|\Psi\rangle$ using $\hat{P}_{1}$:
\begin{equation}
|\Phi\rangle \equiv \mathcal{N}\hat{P}_1|\Psi\rangle=|\phi_{\perp\bar{1}}\rangle|\phi_1^{(2)}\rangle\dots|\phi_1^{(m)}\rangle,
\end{equation}
where $\mathcal{N}$ is a proper normalization factor so that $\langle\Phi|\Phi\rangle=1$.
Then one immediately finds
\begin{equation}\label{en0}
S(\mathrm{Tr}^{(2)}|\Phi\rangle\langle\Phi|)=0.
\end{equation}

On the other hand, since $|\Psi\rangle=|\psi^{(1)}\rangle|\psi^{(2)}\rangle\dots|\psi^{(n)}\rangle$ and the projector $\hat{P}_{1}$ only acts on $|\psi^{(1)}\rangle\dots|\psi^{(l)}\rangle$, $|\Phi\rangle$ can also be written as
\begin{equation}
|\Phi\rangle = \mathcal{N}\hat{P}_1|\Psi\rangle=(\mathcal{N}\hat{P}_1|\psi^{(1)}\rangle\dots|\psi^{(l)}\rangle)|\psi^{(l+1)}\rangle\dots|\psi^{(n)}\rangle.
\end{equation}
So
\begin{equation}\label{EEPhi}
S(\mathrm{Tr}^{(2)}|\Phi\rangle\langle\Phi|)=S(\mathcal{N}\hat{P}_1|\psi^{(1)}\rangle\dots|\psi^{(l)}\rangle\langle\psi^{(l)}|\dots\langle\psi^{(1)}|\hat{P}_1\mathcal{N})+S(\rho_2)=S(\rho_2).
\end{equation}
Combining Eqs. (\ref{Spsi1}), (\ref{en0}), (\ref{EEPhi}), we obtain
\begin{equation}
S(\mathrm{Tr}^{(2)}|\Psi\rangle\langle\Psi|)=0,
\end{equation}
which contradicts Eq. (\ref{EEB}).             \hfill $\Box$

\textit{Extended tridecomposition uniqueness theorem.} \\
If a state is a semi-GHZ state w.r.t. PAT.A: $|\Psi\rangle = \sum_{i=1}^{\bar{i}} \alpha_i |\psi_i^{(1)}\rangle|\psi_i^{(2)}\rangle\dots|\psi_i^{(n)}\rangle$ (denoted as Form A), then within the comparable set of PAT.A, the state {\textit{can not be}} a semi-GHZ state w.r.t. a different partition, say PAT.B: $\sum_{j=1}^{\bar{j}} \kappa_j |\phi_j^{(1)}\rangle|\phi_j^{(2)}\rangle\dots|\phi_j^{(m)}\rangle$ (denoted as Form B), unless $|\psi_i^{(1)}\rangle|\psi_i^{(2)}\rangle\dots|\psi_i^{(n)}\rangle=|\phi_i^{(1)}\rangle|\phi_i^{(2)}\rangle\dots|\phi_i^{(m)}\rangle$ for every $i$ (up to a phase), with a proper ordering of the expanding bases.

The theorem is proved in two steps: first, single out one component of Form A properly, which generates a state being a factorized state w.r.t. PAT.A, meanwhile being an entangled state w.r.t. PAT.B, and apply the lemma on this new state; second, taking the other way around then leads to the theorem.

\textit{Proof.} Since there are two fractions of PAT.B satisfying $\xi_{j_1}^{B\rightarrow A} \bigcap \xi_{j_2}^{B\rightarrow A}=\O$ and $\xi_{j_1}^{B\rightarrow A} \bigcup \xi_{j_2}^{B\rightarrow A}\subsetneq \{A\}$, without losing generality, we take $j_1=1$ and $j_2=2$, and $\xi_{1}^{B\rightarrow A}=\{H_A^{(1)}, H_A^{(2)}, \dots H_A^{(l)}\}$ and $\xi_{2}^{B\rightarrow A}=\{H_A^{(l+1)}, H_A^{(l+2)}, \dots H_A^{(l+l')}\}$. Then we have $\xi_{l+l'+1}^{A\rightarrow B}\subseteq\{H_B^{(3)}, \dots H_B^{(m)}\}$.

Now we try to single out a component of Form A by properly using the states in the subspace ${H_A^{(l+l'+1)}}$. $|\psi_i^{(l+l'+1)}\rangle$ appearing in Form A of $|\Psi\rangle$ are independent, but may not be complete. If they are not complete, then we introduce more independent states $|\psi_i^{(l+l'+1)}\rangle$ with $i$ runs beyond $\bar{i}$, such that the states $|\psi_i^{(l+l'+1)}\rangle$ ($i=1,2,\cdots,d_{H_A^{(l+l'+1)}}$) form an independent and complete set, where $d_{H_A^{(l+l'+1)}}$ is the dimension of the subspace ${H_A^{(l+l'+1)}}$. So noticing that PAT.B is comparable to PAT.A, if $|\Psi\rangle = \sum_{j=1}^{\bar{j}} \kappa_j |\phi_j^{(1)}\rangle|\phi_j^{(2)}\rangle\dots|\phi_j^{(m)}\rangle$, then each state $|\phi_j^{(3)}\rangle\dots|\phi_j^{(m)}\rangle$ ($j=1,\dots,\bar{j}$) can be written as $\sum_{k=1}^{d_{H_A^{(l+l'+1)}}} \gamma_k^j |\psi_k^{(l+l'+1)}\rangle|\bar{\psi}_k^j\rangle$. $|\bar{\psi}_k^j\rangle$ may not be orthogonal or independent. Then we have
\begin{equation}
|\Psi\rangle = \sum_{i=1}^{\bar{i}} \alpha_i |\psi_i^{(1)}\rangle|\psi_i^{(2)}\rangle\dots|\psi_i^{(n)}\rangle=\sum_j^{\bar{j}} \kappa_j |\phi_j^{(1)}\rangle|\phi_j^{(2)}\rangle\left(\sum_k\gamma_k^j|\psi_k^{(l+l'+1)}\rangle|\bar{\psi}_k^j\rangle\right).
\end{equation}
For every $|\psi_i^{(l+l'+1)}\rangle$, one can find at least one state orthogonal to all the states $|\psi_k^{(l+l'+1)}\rangle$ with $k\ne i$ but $|\psi_i^{(l+l'+1)}\rangle$ itself. Denoting this state as $|\psi'_i\rangle$, we have, for $i>\bar{i}$,
\begin{equation}
\langle\psi'_i|\Psi\rangle = 0=\sum_j^{\bar{j}} \kappa_j |\phi_j^{(1)}\rangle|\phi_j^{(2)}\rangle\gamma_i^j\langle\psi'_i|\psi_i^{(l+l'+1)}\rangle|\bar{\psi}_i^j\rangle.
\end{equation}
Since $|\phi_j^{(1)}\rangle$ are independent and all the $\kappa_j$'s are non-zero,
so we have $\gamma_i^j=0$ for $i>\bar{i}$ and any $j$.

For $i\le \bar{i}$, we have
\begin{equation}\label{comp}
\langle\psi'_i|\Psi\rangle =  (\alpha_i \langle\psi'_i|\psi_i^{(l+l'+1)}\rangle)|\psi_i^{(1)}\rangle\dots|\psi_i^{(l+l')}\rangle|\psi_i^{(l+l'+2)}\rangle\dots|\psi_i^{(n)}\rangle=\sum_j^{\bar{j}} (\kappa_j \gamma_i^j \langle\psi'_i|\psi_i^{(l+l'+1)}\rangle) |\phi_j^{(1)}\rangle|\phi_j^{(2)}\rangle|\bar{\psi}_i^j\rangle.
\end{equation}
Then a component of Form A has been properly singled out, on which the lemma presented before is applicable.

According to the lemma, the only possibility for Eq. (\ref{comp}) to hold is that only one of the $\gamma_i^j$'s (for a given $i$) is non-zero. Suppose $\gamma_i^{j_i}$ is non-zero, then we have
\begin{equation}
\langle\psi'_i|\Psi\rangle =  (\alpha_i \langle\psi'_i|\psi_i^{(l+l'+1)}\rangle)|\psi_i^{(1)}\rangle\dots|\psi_i^{(l+l')}\rangle|\psi_i^{(l+l'+2)}\rangle\dots|\psi_i^{(n)}\rangle= (\kappa_{j_i}\gamma_i^{j_i}\langle\psi'_i|\psi_i^{(l+l'+1)}\rangle) |\phi_{j_i}^{(1)}\rangle|\phi_{j_i}^{(2)}\rangle|\bar{\psi}_i^{j_i}\rangle,
\end{equation}
for every $i\le \bar{i}$.
It is important to notice that $\gamma_i^{j\ne j_i}=0$. For a given $i\le\bar{i}$, there exists only one $j=j_i$ such that $\gamma_i^j\ne 0$. In general, for a given $j$ there could be more than one $i$ such that $\gamma_i^j\ne0$ while for all the other $i$'s $\gamma_i^j=0$. This is happening when some $j_i$'s are equal, say $j_{i_a}=j_{i_b}=\dots=\tilde{j}$, then we may write (denoting $W\equiv \{i_a,i_b,\dots\}$)
\begin{equation}
\sum_{i\in W}\frac{|\psi_i^{(l+l'+1)}\rangle\langle\psi'_i|}{\langle\psi'_i|\psi_i^{(l+l'+1)}\rangle}|\Psi\rangle =  \sum_{i\in W}\alpha_i |\psi_i^{(1)}\rangle\dots|\psi_i^{(n)}\rangle= \kappa_{\tilde{j}} |\phi_{\tilde{j}}^{(1)}\rangle|\phi_{\tilde{j}}^{(2)}\rangle\sum_{i\in W}\gamma_i^{\tilde{j}}|\psi_i^{(l+l'+1)}\rangle|\bar{\psi}_i^{\tilde{j}}\rangle.
\end{equation}
Since for this $\tilde{j}$, all the $\gamma_i^{\tilde{j}}=0$ for $i \notin W$, $i$ can run through all legal values, so
\begin{equation}
\sum_{i\in W}\alpha_i |\psi_i^{(1)}\rangle\dots|\psi_i^{(n)}\rangle= \kappa_{\tilde{j}} |\phi_{\tilde{j}}^{(1)}\rangle|\phi_{\tilde{j}}^{(2)}\rangle|\phi_{\tilde{j}}^{(3)}\rangle\dots|\phi_{\tilde{j}}^{(m)}\rangle.
\end{equation}
Since PAT.A is also comparable to PAT.B, according to the lemma, this equation can not hold unless $W$ has only one element. Hence, the only possibility is that for a given $j$ there is only one $i$ such that $\gamma_i^j\ne0$.

To sum up, for every $i\le \bar{i}$ there is only one $j$ such that $\gamma_i^j\ne0$, and for every $j$ there is only one $i$ such that $\gamma_i^j\ne0$. So by rearrange the order we can write $\gamma_i^j\propto\delta_{ij}$ and $\bar{i}=\bar{j}$, then we have
\begin{equation}
\alpha_i |\psi_i^{(1)}\rangle\dots|\psi_i^{(n)}\rangle= \kappa_{i} |\phi_{i}^{(1)}\rangle\dots|\phi_{i}^{(m)}\rangle,
\end{equation}
and $|\alpha_i|=|\kappa_i|$ for every $i\le \bar{i}$. This ends the proof.            \hfill $\Box$

\section{Concluding remarks}\label{Sec5}
In this work, we investigated the problem about the ``uniqueness" of the redundantly recorded observable (of the system of interest) for a given state. After decoherence, the classical observable of the system leaves redundant imprints in the environment, and the information is recorded in disjoint fragments of the environment, which identifies a particular partition of the environment. In general, it is possible for two non-commuting observables to leave redundant imprints simultaneously, if they are associated with two different partitions. Mathematically, any partition is allowed, but not all the partitions have physical significance. So among all the possible partitions of the environment, which partitions are physically meaningful? We need objective criteria to settle this problem, so that we can ensure that only classical observables can proliferate redundantly in the environment. The extended tridecomposition uniqueness theorem proved here combined with the distinguishing of local and nonlocal observers provides one such criterion for a special situation. Although, there are other complicated situations in general, our arguments may help to gain some physical insights. This type of criterion has also been proposed in Ref.~\cite{Loc}, which suggests considering the tensor structure associated with spatial locality and the redundancy of the local records of observables to constrain possible entanglement structures of the state.

The ambiguity we presented in Introduction is not common in dynamical decoherence models. However, for quantum Darwinism to be a fundamental mechanism to explain the emergence of the objectivity of classical properties, it should resolve this ambiguity even if it may occur only in rare conditions, which calls for the present study.

We did not mention how to deal with identical particles, while the constituents of the environment may be identical particles of a few types. To deal with identical particles, we need to turn to quantum field theories. The present form of the theorem cannot be applied to continuum quantum field theories, but can be applied to lattice quantum field theories. We may consider each elementary system to be a state at each spatial point, such as using $|0\rangle_x$, $|1\rangle_x$ and $|2\rangle_x$ to represent the states with 0, 1, and 2 particles (of a particular type) at point $x$ respectively. Then the arguments in the main text are still valid. Currently, it is not clear whether the extension to continuum space-time is straightforward or not, and this problem deserves further efforts in the future.
At last, we expect the uniqueness theorem we proved here is also useful in quantum multipartite entanglement studies.

 \appendix
  \renewcommand{\appendixname}{Appendix}

\section*{Acknowledgments}
We thank C. Jess Riedel for valuable discussions. This work was supported by Natural Science Foundation of China (NSFC) under Grant No. 12047569.

\end{document}